\documentclass[letterpaper, 10 pt, conference]{ieeeconf}  

\IEEEoverridecommandlockouts                              

\title{\LARGE \bf
Help or Hindrance: Understanding the Impact of Robot Communication in Action Teams
}

\author{Tauhid Tanjim$^{1}$, Jonathan St. George$^{2}$, Kevin Ching$^{2}$, and Angelique Taylor$^{1}$
\thanks{*This material was supported by the National Science Foundation under Grant No. IIS-2423127.}
\thanks{$^{1}$T. Tanjim and A. Taylor are with the Department of Information Science at Cornell University, Ithaca, NY 14850, USA {\tt\small\{tt485, amt298\}@cornell.edu}
        }%
\thanks{$^{2}$J. St. George and K. Ching is with Weill Cornell Medicine, Cornell University, New York, NY 10065, USA
        {\tt\small\{jos7007, kec9012\}@med.cornell.edu}}%
}

\usepackage{graphicx}
\usepackage{colortbl}
\usepackage{hyperref}
\usepackage{balance}
\usepackage{flushend}
\usepackage{dblfloatfix}
\usepackage{multirow}
\usepackage{comment}
\usepackage{threeparttable}
\usepackage[normalem]{ulem}

\begin{document}

\maketitle
\thispagestyle{empty}
\pagestyle{empty}

\begin{abstract}

The human-robot interaction (HRI) field has recognized the importance of enabling robots to interact with teams.
Human teams rely on effective communication for successful collaboration in time-sensitive environments.
Robots can play a role in enhancing team coordination through real-time assistance. 
Despite significant progress in human-robot teaming research, there remains an essential gap in how robots can effectively communicate with action teams using multimodal interaction cues in time-sensitive environments.
This study addresses this knowledge gap in an experimental in-lab study to investigate how multimodal robot communication in action teams affects workload and human perception of robots.
We explore team collaboration in a medical training scenario where a robotic crash cart (RCC) provides verbal and non-verbal cues to help users remember to perform iterative tasks and search for supplies.
Our findings show that verbal cues for object search tasks and visual cues for task reminders reduce team workload and increase perceived ease of use and perceived usefulness more effectively than a robot with no feedback.
Our work contributes to multimodal interaction research in the HRI field, highlighting the need for more human-robot teaming research to understand best practices for integrating collaborative robots in time-sensitive environments such as in hospitals, search and rescue, and manufacturing applications.

\end{abstract}

\section{INTRODUCTION}

Human-robot teaming is an active area of research in the human-robot interaction (HRI) field.
As the HRI field moves from one-on-one interaction to human-robot team interaction, the field has recognized the importance of designing robots that engage with multiple people, as it reflects situations that robots encounter in real-world settings \cite{parashar2019taxonomy,iqbal2016movement,taylor2019coordinating}.
This shift involves endowing robots with the capability to understand the needs and intentions of multiple people, reflecting a complex interaction schema.
Effective team communication is crucial to ensure that information flows to the right person and robot at the right time-- in other words, to coordinate the actions of each team member in synchrony \cite{iqbal2015joint}.

The HRI field has investigated how robots can effectively communicate during collaborations with human teams, building on knowledge from human-human interactions in cognitive science \cite{tomasello2005understanding}, psychology \cite{wheatley2024emerging}, and communication fields \cite{clark1991grounding}.
HRI researchers have made significant progress toward understanding how robots can communicate with human teams using navigation \cite{dautzenberg2024follow}, light-based interaction \cite{bacula2023integrating}, and conversational dialogue \cite{tennent2019micbot}, among others.
However, when robots enter time-critical environments, moving from social teams to \textit{action teams}, humans often move dynamically, communicate simultaneously, causing loud environmental noise, and require quick access to information and supplies to perform collaborative tasks effectively.
Thus, robots must be designed to use effective verbal and non-verbal communication, similar to humans, to deliver timely information and supplies. 

Building on knowledge of human-human communication, HRI has explored how robots can communicate with teams verbally and non-verbally.
Verbal communication is the use of natural language via speech from one person to another.
This form of communication is often direct, explicitly stating one's intentions and needs.
Non-verbal communication includes the use of gestures \cite{riek2010cooperative}, eye gaze \cite{admoni2017social}, backchanneling \cite{bliek2020can}, proxemics \cite{mumm2011human}, and visual cues \cite{fukui2013tansubot}.
Visual cues could include light-based interaction, which is beneficial for passive communication where communication is indirectly given from one team member to another, e.g., turning an outdoor light on to indicate help is needed. 
Thus, robots need to balance what information is communicated (verbal or non-verbal) and how this information is given to teams (e.g., speech-based or light-based interactions).
However, designing verbal and non-verbal interactions for robots in time-sensitive settings is not straightforward.

\begin{figure}[t] 
	\centering 
	\includegraphics[width=1.0\linewidth]{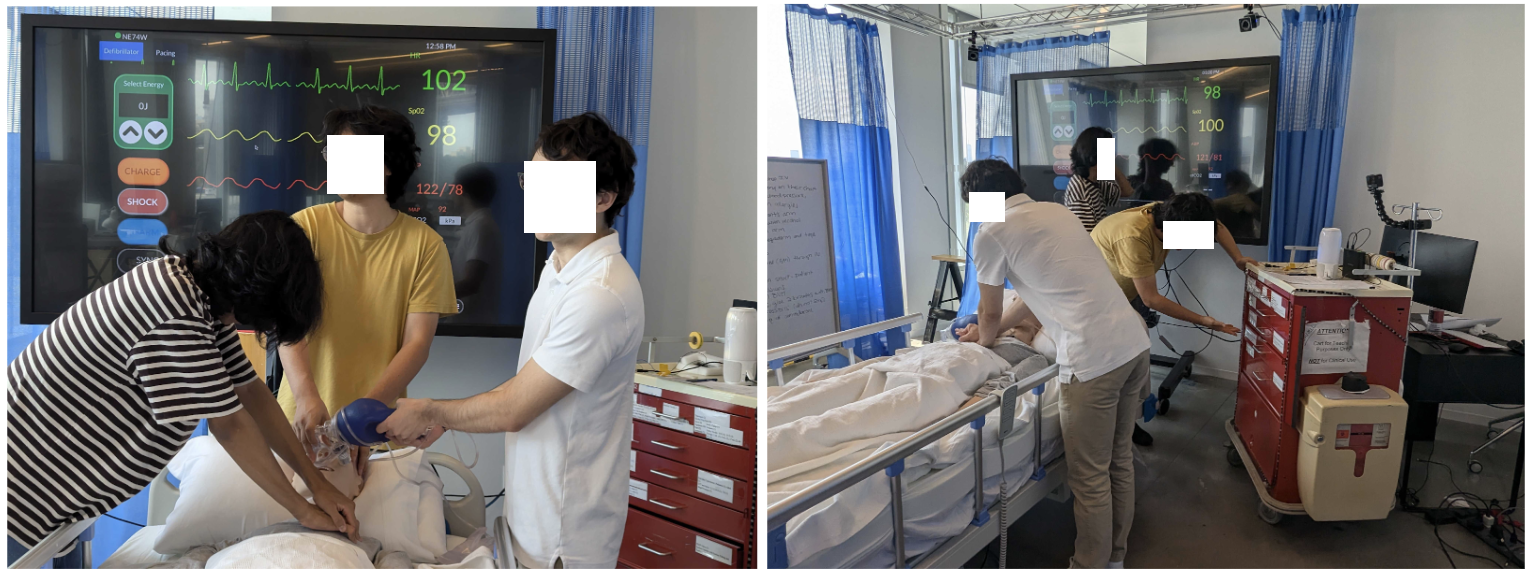} 
	\caption{Multimodal robot communication in human-robot teams.} 
	\label{fig:intro} 
\end{figure}

Prior work has made much progress in understanding how robots can assist teams in time-sensitive environments.
For example, Iqbal et al. \cite{iqbal2016movement} contributed computational models of synchrony to enable robots to take action in group dancing interactions.
Shah et al. proposed goal-driven taxonomies of human-robot team interactions that characterize interaction structures and factors driven by local dynamics (sub-tasks and team collaboration) and contextual factors (task, team, environment) \cite{parashar2019taxonomy}.
Riek et al. \cite{jamshad2024taking} demonstrated the importance of building proactive robots that take initiative in safety-critical environments at the right time to effectively collaborate with action teams.

Despite this significant progress, there is limited knowledge about how robots can effectively communicate in action teams using verbal and non-verbal communication.
Poorly designed robot communication strategies could lead to team mistakes or errors, cause distractions, or reduce team performance instead of improving team collaboration.
Introducing robots into new environments requires well-controlled robot behavior to understand what information they should give and how that information is communicated to action teams at the right time.

Our work addresses these gaps.
We conducted an experimental study to investigate how robot communication in time-sensitive environments affects teamwork.
We explore human-robot teaming in a medical setting where a robot assists a team to treat a patient, a timely and well-suited scenario for robotic assistance \cite{taylor2025rapidly,taylor2019coordinating}.
Building on the work of Taylor et al. \cite{taylor2025rapidly}, we explored how robotic crash carts that store medical supplies and equipment can assist action teams through object search guidance and task reminders, which are common tasks performed by medical teams in hospitals.
Using a Wizard-of-Oz controlled RCC that employs speech-based and light-based communication, we conducted a series of user studies to understand how robots affect team workload, perceived usefulness, and perceived ease of use in medical procedural tasks conducted in in-lab studies.
Thus, we address the following \textbf{research questions}: \textbf{RQ1:} How does verbal communication from robots impact team workload in time-sensitive team collaborations?
\textbf{RQ2:} How does non-verbal communication from robots impact perceived usefulness and perceived ease of use in time-sensitive team collaborations?

We make \textbf{two contributions} to the HRI field. 
First, we conduct a user study that engages participants in a time-sensitive collaborative task with a robot that uses speech- and light-based forms of communication of object search and task reminders.
Our findings indicate a reduction in team workload and increased in perceived usefulness and perceived ease of use when the RCC communicated using speech for object search and light-based interaction for task reminders.  
Second, we discuss how robots operating in time-sensitive environments should adapt their communication style based on the spatial configuration of team members. 
This research reveals the opportunities and risks of human-robot teaming in time-sensitive environments and key lessons learned to push the HRI field toward spatially-aware robot behavior.


\begin{figure}[t] 
	\centering 
	\includegraphics[width=0.5\textwidth]{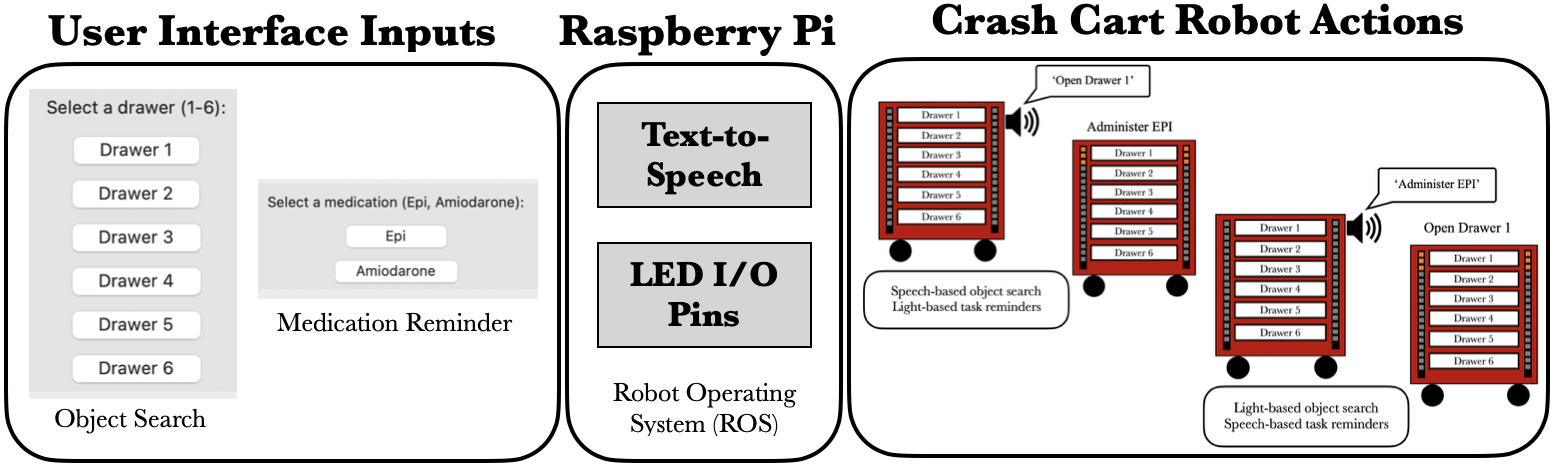} \caption{System diagram of the robotic crash cart illustrating the user interface, processing via Raspberry Pi 4B using ROS, and multimodal outputs through light strips and audio signals.} 
	\label{fig:experiment_diagram} 
\end{figure}

\section{RELATED WORK}

\subsection{Teamwork in Human-Robot Interaction}

Much HRI research has shifted from one-on-one interactions to human-robot interactions in teams \cite{taylor2022regroup, luber2013multi, parashar2019taxonomy}.
Research on action teams in HRI has shown progress towards robots that effectively support and work alongside human teams \cite{ parashar2019taxonomy, sarker2024cohrt, yasar2022robots}.
For example, robots can collaborate with teams using communication cues such as gestures, speech, and graphical user interfaces. Prior work has also explored utilizing goal-oriented task management \cite{angleraud2021coordinating, nikolaidis2015improved},  explainable robot behavior \cite{paleja2021utility, sanneman2020trust}, aligning robot behavior with team values to enhance trust and perceived performance \cite{bhat2024evaluating, gombolay2024human, sanneman2023validating}, and proactive support in healthcare and emergency response settings \cite{haripriyan2024human, jamshad2024taking}.
While these insights have informed the design of robots for team interactions, they have not fully addressed the unique challenges of enabling robots to collaborate effectively with teams in time-sensitive interactions.
We address this gap by conducting user studies to empirically evaluate the impact of multimodal communication strategies on workload and human perception of robots in time-sensitive human-robot teaming scenarios.
Our approach focuses on combining verbal and non-verbal feedback to support intuitive task reminders, minimize time spent searching for
supplies, and maintain team situation awareness without disrupting human interaction workflows.

\subsection{Multimodal Interaction in HRI}

A large body of literature has explored multimodal interaction between robots and humans using verbal and non-verbal communication, such as lights \cite{fukui2013tansubot}, sound \cite{song2019designing, pelikan2023designing}, motion \cite{cha2018effects}, and speech \cite{lima2024home}, which have shown potential for improving human-robot interaction.
For example, TansuBot is a cart-based robot that uses LED indicators with visual feedback to guide users in object search tasks \cite{fukui2013tansubot}, and Go to Any Thing (GOAT) used multimodal input for navigation \cite{Chang-RSS-24}. 
A key challenge of deploying robots in highly collaborative teams lies in designing multimodal cues easily interpretable by human collaborators \cite{terziouglu2020designing}.
Ongoing research efforts have explored this challenge in various environments, such as public buses \cite{pelikan2023designing}, households \cite{fukui2013tansubot}, manufacturing industries \cite{papanastasiou2019towards}, and healthcare \cite{addlesee2024multi}.
However, the specific combination of multimodal task-specific interaction in human-robot teaming remains underexplored, particularly in collaborative tasks such as emergency resuscitation codes \cite{taylor2019coordinating, jamshad2024taking, matsumoto2023robot}.
The work by Taylor et al. \cite{taylor2024towards} shows the potential for medical crash cart robots to support team collaboration in hospital settings using multimodal cues.
Thus, we focus on robotic visual cues and speech-based task assistance in human teams.

\begin{figure}[t] 
	\centering 
	\includegraphics[width=0.8\linewidth]{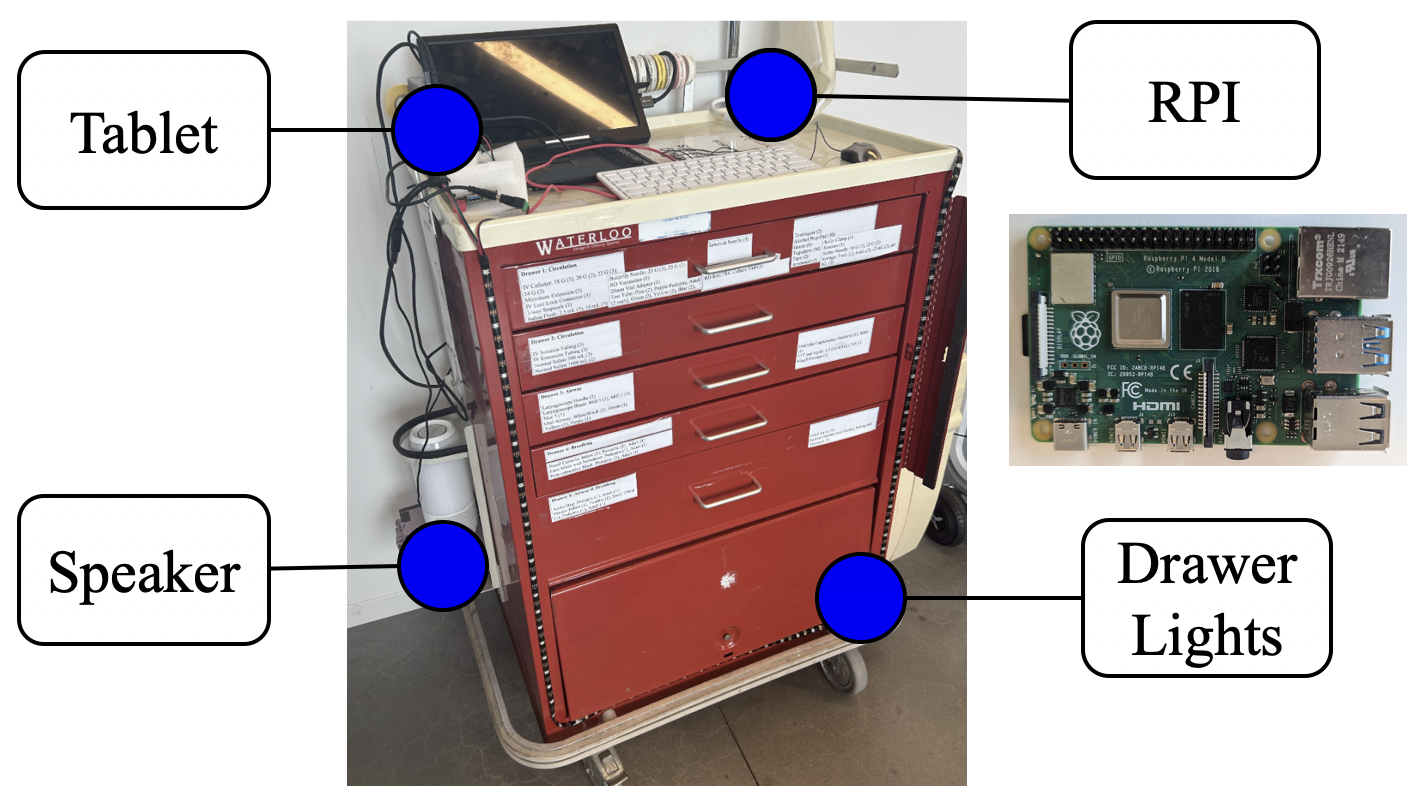} 
\caption{The Robotic Crash Cart (RCC) was operated on a tablet using Wizard-of-Oz control driven by a Raspberry Pi that sent signals to a speaker and LED light strip to communicate with action teams.} 
	\label{fig:Wizard_RCC_Setup} 
\end{figure}

\section{METHODOLOGY}

We conducted a series of pilot studies to investigate use-cases best suited for a cart-based robot in medical settings during medical procedures, which often necessitate effective team collaboration. Based on lessons learned, we conducted a main study with a large sample size to understand the trade-offs of robot mediation using verbal and non-verbal communication in time-sensitive team collaborations.

\subsection{Robotic Platform}

We built a robotic crash cart (RCC) using the Crash Cart Robot Toolkit\footnote{\url{https://github.com/Cornell-Tech-AIRLab/crash_cart_robot_tutorial} \label{github_link}}, released by Taylor et al. \cite{taylor2025rapidly}.
This robot is a 6-drawer, teleoperated platform used during medical training sessions at a local medical school.
The robot is driven by a Raspberry Pi (RPI) 4B running Robot Operating System 2 (ROS2) that sends messages to a tablet, Bluetooth speaker, and LED strip placed around the outer edge of the cart drawers (see Figure \ref{fig:Wizard_RCC_Setup}).
The tablet displays a user interface to Wizard-of-Oz control the robot, connects to the RPI to control when a series of LED lights turn on and off (i.e., LED blinking) for a selected drawer, and transmits an audio signal to the speaker to enable the robot to communicate with teams using speech.

\subsection{Pilot Study} 

We conducted an IRB-approved pilot study to understand what and when robot communication is most effective for robots to recommend task reminders and locations of relevant supplies in a hospital setting.
Taylor et al. \cite{taylor2025rapidly} collected feedback from healthcare workers and found that robotic speech, lights, and alerts may prove useful for RCCs.
Inspired by this work, we conducted iterative pilot studies to ensure that 
1) the study environment reflected a real-world hospital room in a lab setting,
2) participants received effective training to perform a medical procedure using a simulation training scenario provided by our collaborators, medical educators at a local medical school,
3) and users interpret the robot's intended use of sounds, lights, and speech recommendations for object search and task reminder tasks.

\textbf{Design of Robot Communication Modalities and Study Task:} We started by accessing individual robot sounds, LED blinking, and speech for task reminders and object search tasks. 
As shown in prior work, we found that multimodal communication was more effective than unimodal communication \cite{axelsson2022multimodal}. 
Thus, we focused on multimodal robot communication in the remaining pilot studies.
Through our active collaboration with a local medical school, we consulted with two medical educations to determine a set of fixed and adjustable time intervals to Wizard-of-Oz control robot recommendations. 
We found that reminders for administering medications and object searches are tasks that often cause delays during human-cart interactions; thus, we designed robot communication strategies for these tasks.
We started with Advanced Life Support (ALS), which consisted of standalone and iterative tasks conducted at fixed time intervals. The steps of ALS include: 1) initial patient assessment of pulse, chest rise, and breathing; 2) cardiopulmonary resuscitation (CPR) and placement of bag valve mask ventilation; 3) checking patient medical history and setting up for intravenous line (IV) administration; 4) administering medication; and 5) repeating prior steps as needed.
A wizard was situated behind the robot in the study room.

\textbf{Participants:} We recruited 19 participants (3-4 participants per session) to engage in 6 user studies using flyers placed around a university campus in the global north. 
Participants ages ranged from 19 to 31 (M = 26, STD = 3.6) with 6 females and 8 males. 
4 participants were practicing healthcare workers.
Participants rated their familiarity with robots from (little familiarity) to 5 (very familiar), with a familiarity ranging from 1 to 5 (M=3.6, STD = 1.02).
All participants were compensated with an \$18 Amazon gift card.
We assigned participants to study sessions using convenience sampling.

\textbf{Study Design and Experimental Testbed:} We conducted a within-subjects study with three study conditions: C1) LED blinking for object search guidance, C2) LED blinking and speech for task reminders and object search guidance, and C3) control group (no feedback).
The order of conditions was randomized, and study participants were instructed to rotate tasks throughout the study to ensure they all engaged with the robot at least once.
We randomly placed items in different drawers between conditions to minimize carryover effects \cite{macfie1989designs}. 
We conducted the study in a lab space designed to resemble a realistic medical training room. We used a patient bed, a static patient manikin, RCC, whiteboards with task instructions, and a large monitor displaying vital signs and emitting beeping loud sounds through the Simpl Patient Monitor application to set up the training room.

\textbf{Wizard Protocol:} A wizard was a member of our research team situated in the lab behind the RCC to control the robot actions during the study. The goal of the wizard was to teleoperate the robot to recommend supplies and provide medication reminders based on participant conversations and task steps. Administering medication is an iterative task; thus, the wizard generated reminders from the robot at fixed time intervals.
Using the tablet-based user interface, the wizard performed the following tasks: 
\begin{itemize}
\item Turn LED lights on and off three times for the drawer that contains supplies for a task or turn all drawer lights on and off three times for medication reminders.
\item Use text-to-speech software to send a dialogue speech prompt to the team. The robot said, 'Kindly open drawer 1' when the required supplies were in the first drawer, and the robot said, 'Administer EPI' to remind participants to administer medication. 
\end{itemize}

\textbf{Study Procedure:} Participants engaged in one-hour study sessions that involved 1) the experimenter introduced the study goals to participants and trained them by demonstrating the study tasks, 2) participants engaged in three five-minute ALS procedures to capture the time-sensitive nature of medical tasks, 3) participants completing self-report measures after each session, and 4) the experimenter facilitated a debrief discussion to collect open-ended feedback about interactions with the robot.

\textbf{Data Collection and Analysis:} We collected video and self-report measures to understand how robot speech and light-based feedback affected team workload. We recorded video from 3-4 overhead cameras and 1 camera onboard the RCC. We measured workload using the NASA Task Load Index (NASA-TLX), a 6-item 7-point Likert scale measure of mental demand, physical demand, temporal demand, performance, effort, and frustration \cite{hart1988development}. We analyzed workload using descriptive statistics. We transcribed the debrief session audio data and analyzed it using grounded theory \cite{glaser2017discovery}.

\begin{figure}[t] 
	\centering 
	\includegraphics[width=1.0\linewidth]{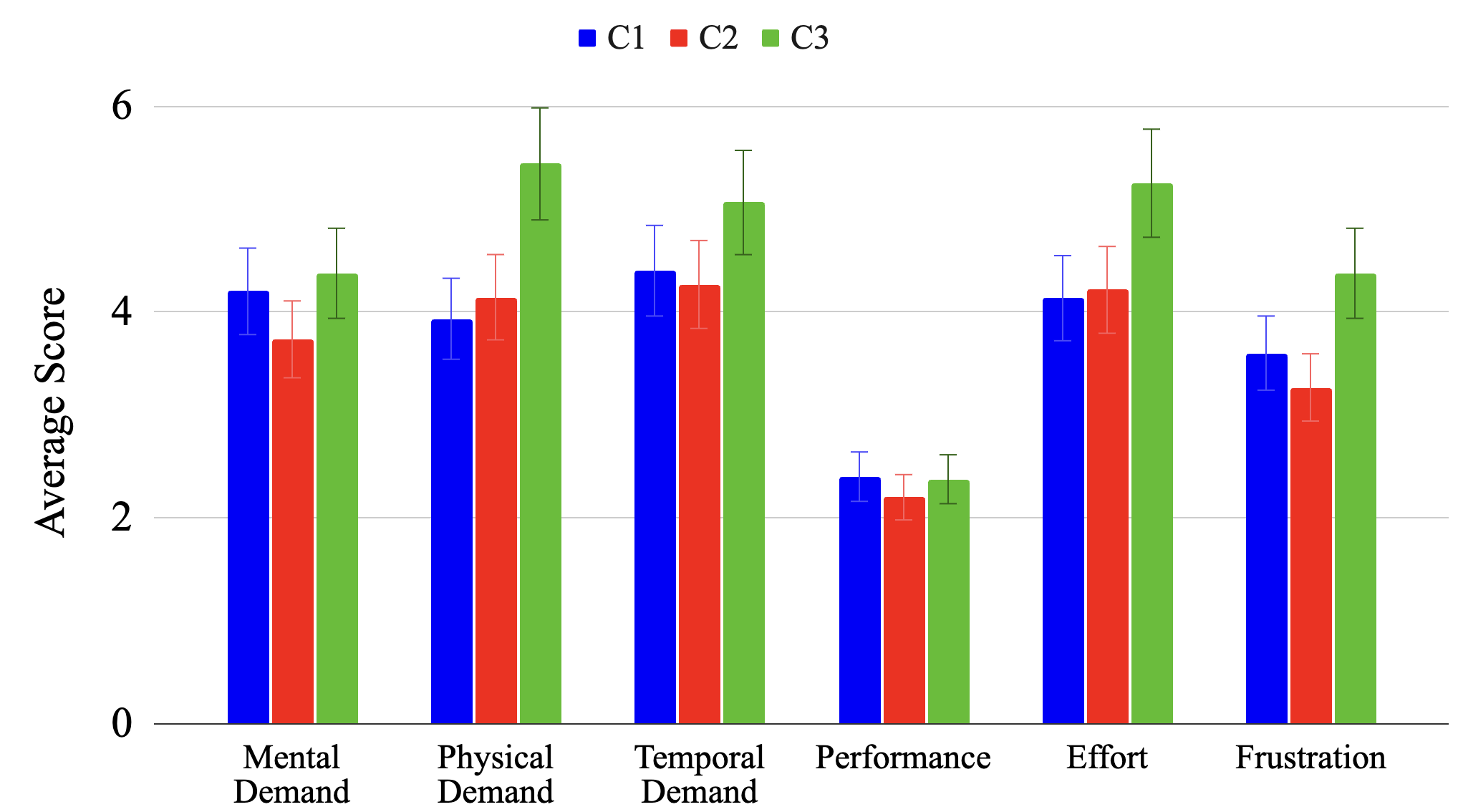} 	\caption{Pilot study NASA-TLX workload scores.} 
	\label{fig:pilot_results} 
\end{figure}

\textbf{Pilot Results:} Our results indicate that C3 achieved in the maximum workload in terms of mental, temporal, effort, and frustration demand (see Figure \ref{fig:pilot_results}). Participants reported the minimum workload in terms of physical and effort demand in C1.
C2 achieved the lowest average score for demand, performance demand, temporal demand, and frustration.

We started pilot studies using lights, speech, and sounds as communication modalities to enable the RCC to assist in team collaborators.
We learned in our initial pilot study debriefs that sounds could be useful for quick, iterative tasks such as CPR compressions.
However, participants found alerts were difficult to interpret for object search and medical reminder tasks-- that is, participants did not understand the robot task recommendation. 
As a result, we focused on speech- and light-based communication in the remainder of pilot studies.
We found that the robot speech utterances (e.g., ‘Kindly open drawer 1' or `EPI is in drawer 1’) were too long. Participants often shuffled through drawers waiting for the robot to finish its phrase. Medication reminders were short (`Administer EPI') and served as a good example for short speech phrases.
Lastly, participants confirmed that LED blinking effectively captured their attention during object search and task reminders.

\subsection{Main Study} 

We conducted an IRB-approved main study to understand the impact of robot communication during team collaboration using lessons learned from the pilot study. 

\textbf{Participants:} We recruited 84 participants using convenience sampling by posting flyers around a university campus and on list-serves.
Participant ages ranged from 21 to 79 (M = 25.9, STD = 6.76), with 47 female, 34 male, one non-binary, and two who preferred not to disclose their gender. 
We conducted 26 studies with 3--4 participants per study session.
39 participants had a clinical background: 3 participants had 3--5 years, 3 participants had 1--2 years, and 33 participants had 0--6 months.
Participants' familiarity with robots ranged from 1 to 5 (M=2.5, STD=1.1).
All participants were compensated with an \$18 Amazon gift card.

\textbf{Study Design, Study Task, Experimental Testbed, Wizard Protocol, Procedure:} We used the same study task, experimental testbed, wizard protocol, and procedure from the pilot study in the main study. We changed the study conditions to understand the trade-offs between speech- and light-based communication during for both object search and medication reminders in time-sensitive team collaborations.
Thus, we used a within-subjects study design with three conditions: C1) LED blinking for object search guidance and speech for medication reminders, C2) LED blinking for task reminders and speech for object search, and C3) control group.
To address challenges in pilot studies, we shortened the robot's speech utterances for object search to fixed prompts (e.g., ‘Drawer 1’) and placed numbered labels on the cart drawers to prevent participants from counting the drawers, potentially adding to mental demand.

\textbf{Data Collection and Analysis:} Data collection was consistent with the pilot study with one exception.
In addition to administering the NASA-TLX survey, we also administered the 12-item 7-point Likert scale Technology Acceptance Model (TAM) \cite{davis1989technology} to understand participants' perceived usefulness (PU) and perceived ease of use (PEU) of the robot. 
A one-way repeated analysis of variance (ANOVA) test was used to determine whether there is an effect of PEU, PU, and workload on study conditions. 
Following significant effects, we conducted post hoc comparisons using the Bonferroni Correction test to examine pairwise differences between conditions (C1 vs C2, C2 vs C3, and C1 vs C3).
We applied Greenhouse-Geisser corrections for sphericity violations, while other minor assumption violations were considered negligible given our sample size. 

\section{Main Study Results}

\begin{figure}[t] 
	\centering 
	\includegraphics[width=0.5\textwidth]{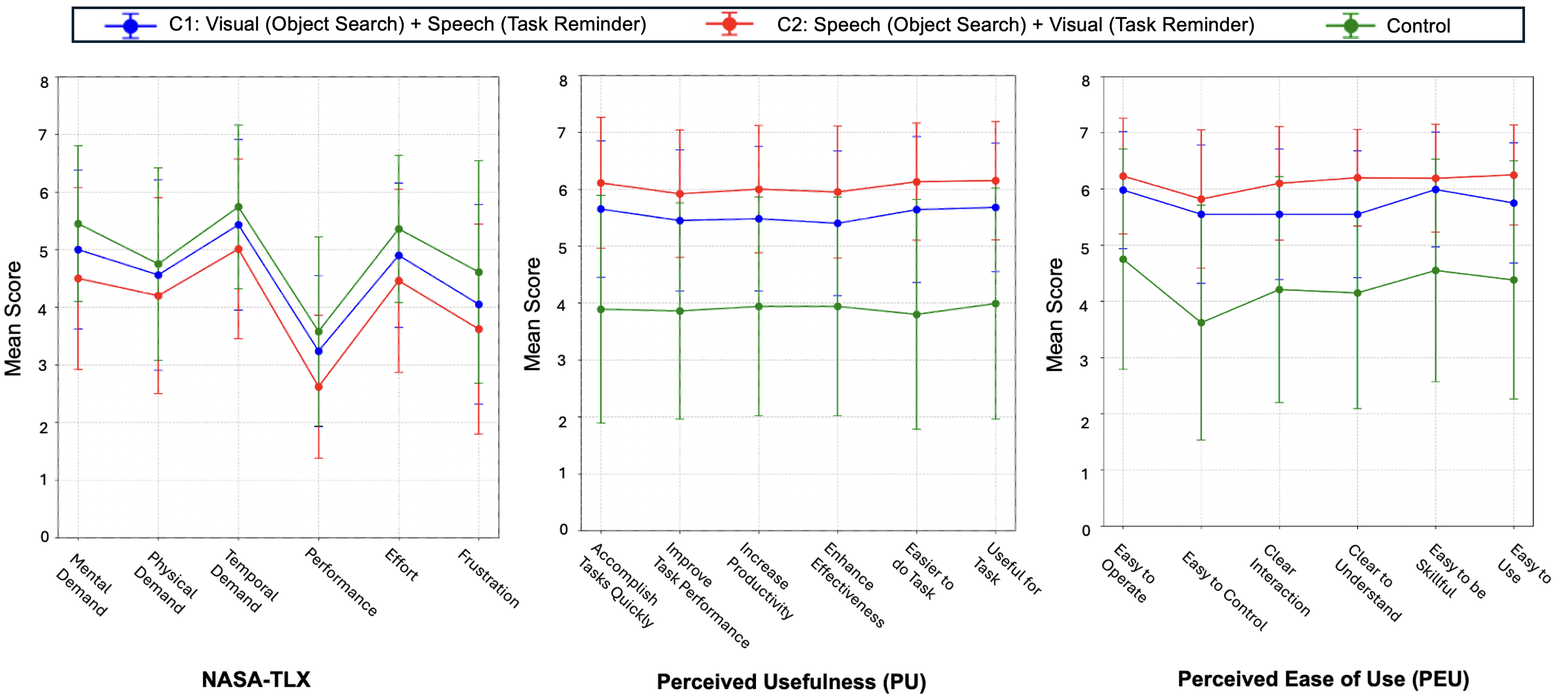} 	\caption{Main study results: NASA-TLX (left), Perceived Usefulness (middle) and Perceived Ease of Use (right).}
	\label{fig:pu_peu_results} 
\end{figure}

Figure \ref{fig:pu_peu_results} shows the statistical analysis results for all study conditions and measures.

\subsection{Perceived Usefulness Results}

We found significant main effects of all PU measures including accomplishing tasks quickly, $F(2, 166) = 68.470$, $p < .001$, partial $\eta^2 = .452$; improving task performance $F(2, 166) = 59.985$, $p < .001$, partial $\eta^2 = .420$; increasing productivity, $F(2, 166) = 55.310$, $p < .001$, partial $\eta^2 = .400$; enhancing effectiveness, $F(2, 166) = 51.045$, $p < .001$, partial $\eta^2 = .381$; easier to perform task, $F(2, 166) = 74.135, p < .001$; and the perceived utility for tasks $F(2, 166) = 62.585$, $p < .001$, partial $\eta^2 = .430$. The post hoc tests revealed a significantly higher PU in C2 than C1 and C3 which suggests that participants perceived the RCC with verbal object search and visual reminders as significantly more useful than in C1 and C3. Additionally, C1 was significantly higher than C3, which indicates that participants found visual object search and verbal reminders more useful than the traditional cart.

\subsection{Perceived Ease of Use Results}
We found a significant difference for ease of operating the RCC, $F(2, 166) = 29.424$, $p < .001$, partial $\eta^2 = .262$; ease of controlling the RCC, $F(2, 166) = 58.187$, $p < .001$, partial $\eta^2 = .412$; and clarity of interaction, $F(2, 166) = 47.498$, $p < .001$, partial $\eta^2 = .364$; clarity of understanding the RCC, $F(2, 166) = 56.005$, $p < .001$, partial $\eta^2 = .403$; ease of becoming skillful, $F(2, 166) = 46.524$, $p < .001$, partial $\eta^2 = .359$; and overall ease of use for task completion, $F(2, 166) = 46.180$, $p < .001$, partial $\eta^2 = .357$.
In the post hoc tests, we discovered that ease of use was significantly higher in C2 than C1 and C3. This suggests that 
verbal object search and visual reminders were significantly easier to use in C2 than C1 and C3. 
Participants rated C1 significantly higher than C3, which indicates that visual object search with verbal reminders were easier to use than the traditional crash cart.

\subsection{Workload Results}

We found significant differences for mental demand, $F(2, 166) = 15.993$, $p < .001$, partial $\eta^2 = .162$; physical demand, $F(2, 166) = 7.264$, $p = .009$, partial $\eta^2 = .080$; and temporal demand, $F(2, 166) = 7.204$, $p = .001$, partial $\eta^2 = .080$. Additionally, we observed significant effects for performance, $F(2, 166) = 11.508$, $p < .001$, partial $\eta^2 = .122$; effort, $F(2, 166) = 14.795$, $p < .001$, partial $\eta^2 = .151$; and frustration, $F(2, 166) = 12.460$, $p < .001$, partial $\eta^2 = .131$.
We discovered in the post hoc tests that workload was significantly lower in C2 compared to C1 and C3. Thus, verbal object search and visual reminders were significantly less demanding workload in C2 than in C1 and C3.
Participants rated C1 significantly lower than C3 which shows that visual object search and verbal reminders were less demanding than the traditional crash cart.

\section{DISCUSSION}

Our results revealed that robots can effectively reduce workload, improve perceived usefulness and ease of use in time-sensitive team collaborations compared to no assistance.
Furthermore, we learned that there are many factors that impact the best use of robot communication in action teams.

\textit{Impact of Spatial Dynamics}: Our study highlights the importance of aligning robot communication with action teams' spatial dynamics-that is, their position and orientation throughout collaboration particularly in small spaces, consistent with prior literature on proxemics in HRI \cite{mumm2011human}. 
Unique to time-sensitive team collaborations, robots need to quickly communicate interpretable feedback to action teams, adjusting itsorientation to those more relevant to its feedback.
For instance, object search tasks are most relevant to a user retrieving supplies, while medication reminders are useful for the entire team.
On the other hand, if no users are near the robot, action teams risk missing the visual cues from the robot.
These findings are a design tension with the failure taxonomies from Taylor et al. \cite{taylor2025rapidly}, who found that robot movement during team collaborations using cart-based robots could cause suggestive, obstructive, and distractive failures. 
Furthermore, robot speech served as a direct form of communication that is beneficial irrespective of team member spatial dynamics.
However, spaces in some applications are inherently loud and small; therefore, this necessitates careful design of speech-based capabilities to avoid increasing alarm fatigue (i.e., Emergency Medicine) \cite{fujita2020customizing}.
Thus, the alignment of robot communication modalities, timing, and orientation is crucial to ensure new errors are not introduced.

\textit{Impact of Robot Embodiment}: Consistent with prior work, we found that the embodiment set users' expectations of how to communicate with the robot \cite{riek2010cooperative}.
Unique to our study, cart-based robots set clear expectations around equipment retrieval and task reminders; however, our study highlights the need for robots to use consistent communication strategies for specific tasks to avoid misinterpreting intended assistive behaviors.
Thus, other embodied robots may require different expectation setting before human-robot teaming begins.
As a result, communication modes must be carefully designed to ensure that action teams interpret the robot correctly--leading to helping instead of hindering collaboration.

\section{Conclusion}

In this work, we explored multimodal robot communication during time-sensitive team collaborations. Our findings highlight the nuances of designing effective robot communication to ensure that robots are positively perceived by action teams. 
Our analysis underscores the need for robots to use adaptive communication strategies to account for the spatial dynamics of team members and environmental constraints.
By aligning robot communication timing and multimodal interaction with the team role and needs, robots can better adapt and support team collaboration.
Our work contributes to collaborative robotics research in the HRI community.

\bibliographystyle{IEEEtran}

\begin{thebibliography}{10}
\providecommand{\url}[1]{#1}
\csname url@samestyle\endcsname
\providecommand{\newblock}{\relax}
\providecommand{\bibinfo}[2]{#2}
\providecommand{\BIBentrySTDinterwordspacing}{\spaceskip=0pt\relax}
\providecommand{\BIBentryALTinterwordstretchfactor}{4}
\providecommand{\BIBentryALTinterwordspacing}{\spaceskip=\fontdimen2\font plus
\BIBentryALTinterwordstretchfactor\fontdimen3\font minus \fontdimen4\font\relax}
\providecommand{\BIBforeignlanguage}[2]{{%
\expandafter\ifx\csname l@#1\endcsname\relax
\typeout{** WARNING: IEEEtran.bst: No hyphenation pattern has been}%
\typeout{** loaded for the language `#1'. Using the pattern for}%
\typeout{** the default language instead.}%
\else
\language=\csname l@#1\endcsname
\fi
#2}}
\providecommand{\BIBdecl}{\relax}
\BIBdecl

\bibitem{parashar2019taxonomy}
P.~Parashar, L.~M. Sanneman, J.~A. Shah \emph{et~al.}, ``A taxonomy for characterizing modes of interactions in goal-driven, human-robot teams,'' in \emph{Proc. IEEE/RSJ Intern. Conf. on Intelligent Robots and Systems (IROS)}, 2019, pp. 2213--2220.

\bibitem{iqbal2016movement}
T.~Iqbal, S.~Rack, and L.~D. Riek, ``Movement coordination in human--robot teams: A dynamical systems approach,'' \emph{IEEE Trans. on Robotics}, vol.~32, no.~4, pp. 909--919, 2016.

\bibitem{taylor2019coordinating}
A.~Taylor, H.~R. Lee, A.~Kubota \emph{et~al.}, ``Coordinating clinical teams: Using robots to empower nurses to stop the line,'' \emph{Proc. ACM on Human-Computer Interaction}, vol.~3, no. CSCW, pp. 1--30, 2019.

\bibitem{iqbal2015joint}
T.~Iqbal, M.~J. Gonzales, L.~D. Riek \emph{et~al.}, ``Joint action perception to enable fluent human-robot teamwork,'' in \emph{Proc. IEEE Intern. Symp. Robot Human Interactive Communication (RO-MAN)}, 2015.

\bibitem{tomasello2005understanding}
M.~Tomasello, M.~Carpenter, J.~Call \emph{et~al.}, ``Understanding and sharing intentions: The origins of cultural cognition,'' \emph{Behavioral and brain sciences}, vol.~28, no.~5, pp. 675--691, 2005.

\bibitem{wheatley2024emerging}
T.~Wheatley, M.~A. Thornton, A.~Stolk \emph{et~al.}, ``The emerging science of interacting minds,'' \emph{Perspect. on Psychological Science}, vol.~19, no.~2, pp. 355--373, 2024.

\bibitem{clark1991grounding}
H.~H. Clark and S.~E. Brennan, ``Grounding in communication,'' 1991.

\bibitem{dautzenberg2024follow}
P.~Dautzenberg, S.~Ladwig, and A.~M. Rosenthal-von~der Pütten, ``Follow me: Anthropomorphic appearance and communication impact social perception and joint navigation behavior,'' in \emph{Proc. ACM/IEEE Intern. Conf. Human-Robot Interaction (HRI)}, 2024, pp. 175--183.

\bibitem{bacula2023integrating}
A.~Bacula, J.~Mercer, J.~Berger \emph{et~al.}, ``Integrating robot manufacturer perspectives into legible factory robot light communications,'' \emph{ACM Trans. Human-Robot Interaction}, vol.~12, no.~1, pp. 1--33, 2023.

\bibitem{tennent2019micbot}
H.~Tennent, S.~Shen, and M.~Jung, ``Micbot: A peripheral robotic object to shape conversational dynamics and team performance,'' in \emph{Proc. ACM/IEEE Int. Conf. Human-Robot Interaction (HRI)}, 2019.

\bibitem{riek2010cooperative}
L.~D. Riek, T.-C. Rabinowitch, P.~Bremner \emph{et~al.}, ``Cooperative gestures: Effective signaling for humanoid robots,'' in \emph{Proc. ACM/IEEE Intern. Conf. on Human-Robot Interaction (HRI)}, 2010, pp. 61--68.

\bibitem{admoni2017social}
H.~Admoni and B.~Scassellati, ``Social eye gaze in human-robot interaction: A review,'' \emph{J. Human-Robot Interaction}, 2017.

\bibitem{bliek2020can}
A.~Bliek, S.~Bensch, and T.~Hellström, ``How can a robot trigger human backchanneling?'' in \emph{Proc. IEEE Intern. Conf. Robot and Human Interactive Communication (RO-MAN)}, 2020, pp. 96--103.

\bibitem{mumm2011human}
J.~Mumm and B.~Mutlu, ``Human-robot proxemics: Physical and psychological distancing in human-robot interaction,'' in \emph{Proc. 6th Int. Conf. Human-Robot Interaction (HRI)}, 2011, pp. 331--338.

\bibitem{fukui2013tansubot}
R.~Fukui, T.~Sunakawa, S.~Kousaka \emph{et~al.}, ``Tansubot: A drawer-type storage system for supporting object search with contents' photos and usage histories,'' in \emph{Proc. IEEE/RSJ Int. Conf. on Intelligent Robots and Systems (IROS)}, 2013, pp. 2703--2709.

\bibitem{jamshad2024taking}
R.~Jamshad, A.~Haripriyan, A.~Sonti \emph{et~al.}, ``Taking initiative in human-robot action teams: How proactive robot behaviors affect teamwork,'' in \emph{Proc. 2024 ACM/IEEE Intern. Conf. on Human-Robot Interaction}, 2024, pp. 559--562.

\bibitem{taylor2025rapidly}
A.~Taylor, T.~Tanjim, M.~J. Sack, M.~Hirsch, K.~Cheng, K.~Ching, J.~S. George, T.~Roumen, M.~F. Jung, and H.~R. Lee, ``Rapidly built medical crash cart! lessons learned and impacts on high-stakes team collaboration in the emergency room,'' in \emph{2025 20th ACM/IEEE International Conference on Human-Robot Interaction (HRI)}.\hskip 1em plus 0.5em minus 0.4em\relax IEEE, 2025, pp. 501--510.

\bibitem{taylor2022regroup}
A.~Taylor and L.~D. Riek, ``Regroup: A robot-centric group detection and tracking system,'' in \emph{Proc. 17th ACM/IEEE Intern. Conf. on Human-Robot Interaction (HRI)}, 2022, pp. 412--421.

\bibitem{luber2013multi}
M.~Luber and K.~O. Arras, ``Multi-hypothesis social grouping and tracking for mobile robots.'' in \emph{Robotics: Science and Systems}, 2013.

\bibitem{sarker2024cohrt}
S.~Sarker, H.~N. Green, M.~S. Yasar \emph{et~al.}, ``Cohrt: A collaboration system for human-robot teamwork,'' \emph{arXiv:2410.08504}, 2024.

\bibitem{yasar2022robots}
M.~S. Yasar and T.~Iqbal, ``Robots that can anticipate and learn in human-robot teams,'' in \emph{Proc. 17th ACM/IEEE Intern. Conf. on Human-Robot Interaction (HRI)}, 2022, pp. 1185--1187.

\bibitem{angleraud2021coordinating}
A.~Angleraud, A.~Mehman~Sefat, M.~Netzev \emph{et~al.}, ``Coordinating shared tasks in human-robot collaboration by commands,'' \emph{Frontiers in Robotics and AI}, vol.~8, p. 734548, 2021.

\bibitem{nikolaidis2015improved}
S.~Nikolaidis, P.~Lasota, R.~Ramakrishnan \emph{et~al.}, ``Improved human--robot team performance through cross-training, an approach inspired by human team training practices,'' \emph{Int. J. Robotics Res.}, vol.~34, no.~14, pp. 1711--1730, 2015.

\bibitem{paleja2021utility}
R.~Paleja, M.~Ghuy, N.~Ranawaka~Arachchige \emph{et~al.}, ``The utility of explainable ai in ad hoc human-machine teaming,'' \emph{Adv. Neural Inf. Process. Syst.}, vol.~34, pp. 610--623, 2021.

\bibitem{sanneman2020trust}
L.~Sanneman and J.~A. Shah, ``Trust considerations for explainable robots: A human factors perspective,'' \emph{arXiv:2005.05940}, 2020.

\bibitem{bhat2024evaluating}
S.~Bhat, J.~B. Lyons, C.~Shi, and X.~J. Yang, ``Evaluating the impact of personalized value alignment in human-robot interaction: Insights into trust and team performance outcomes,'' in \emph{Proceedings of ACM/IEEE International Conference on Human-Robot Interaction}, 2024.

\bibitem{gombolay2024human}
M.~Gombolay, ``Human-robot alignment through interactivity and interpretability: Don’t assume a “spherical human”,'' in \emph{Proceedings of the Thirty-Third International Joint Conference on Artificial Intelligence}, 2024, pp. 8523--8528.

\bibitem{sanneman2023validating}
L.~Sanneman and J.~A. Shah, ``Validating metrics for reward alignment in human-autonomy teaming,'' \emph{Computers in Human Behavior}, vol. 146, p. 107809, 2023.

\bibitem{haripriyan2024human}
A.~Haripriyan, R.~Jamshad, P.~Ramaraj \emph{et~al.}, ``Human-robot action teams: A behavioral analysis of team dynamics,'' in \emph{Proc. 33rd IEEE Intern. Conf. on Robot and Human Interactive Communication (ROMAN)}, 2024, pp. 1443--1448.

\bibitem{song2019designing}
S.~Song and S.~Yamada, ``Designing led lights for a robot to communicate gaze,'' \emph{Adv. Robotics}, vol.~33, no. 7-8, pp. 360--368, 2019.

\bibitem{pelikan2023designing}
H.~R. Pelikan and M.~F. Jung, ``Designing robot sound-in-interaction: The case of autonomous public transport shuttle buses,'' in \emph{Proc. 2023 ACM/IEEE Intern. Conf. on Human-Robot Interaction}, 2023.

\bibitem{cha2018effects}
E.~Cha, N.~T. Fitter, Y.~Kim \emph{et~al.}, ``Effects of robot sound on auditory localization in human-robot collaboration,'' in \emph{Proc. 2018 ACM/IEEE Intern. Conf. on Human-Robot Interaction}, 2018, pp. 434--442.

\bibitem{lima2024home}
M.~R. Lima, ``Home integration of conversational robots to enhance ageing and dementia care,'' in \emph{Proc. 2024 ACM/IEEE Intern. Conf. on Human-Robot Interaction}, 2024, pp. 115--117.

\bibitem{Chang-RSS-24}
M.~Chang, T.~Gervet, M.~Khanna \emph{et~al.}, ``{GOAT: GO to Any Thing},'' in \emph{Proc. Robotics: Science and Systems}, Delft, Netherlands, 2024.

\bibitem{terziouglu2020designing}
Y.~Terzioğlu, B.~Mutlu, E.~Şahin \emph{et~al.}, ``Designing social cues for collaborative robots: The role of gaze and breathing in human-robot collaboration,'' in \emph{Proc. 2020 ACM/IEEE Intern. Conf. on Human-Robot Interaction}, 2020, pp. 343--357.

\bibitem{papanastasiou2019towards}
S.~Papanastasiou, N.~Kousi, P.~Karagiannis \emph{et~al.}, ``Towards seamless human robot collaboration: Integrating multimodal interaction,'' \emph{Int. J. Adv. Manufacturing Technology}, vol. 105, pp. 3881--3897, 2019.

\bibitem{addlesee2024multi}
A.~Addlesee, N.~Cherakara, N.~Nelson \emph{et~al.}, ``Multi-party multimodal conversations between patients, their companions, and a social robot in a hospital memory clinic,'' in \emph{Proc. 18th Conf. of the European Chapter of the Assoc. for Computational Linguistics}, 2024.

\bibitem{matsumoto2023robot}
S.~Matsumoto, P.~Ghosh, R.~Jamshad, and L.~D. Riek, ``Robot, uninterrupted: Telemedical robots to mitigate care disruption,'' in \emph{Proceedings of the 2023 ACM/IEEE International Conference on Human-Robot Interaction}, 2023, pp. 495--505.

\bibitem{taylor2024towards}
A.~Taylor, T.~Tanjim, H.~Cao, and H.~R. Lee, ``Towards collaborative crash cart robots that support clinical teamwork,'' in \emph{Proceedings of the 2024 ACM/IEEE International Conference on Human-Robot Interaction}, 2024, pp. 715--724.

\bibitem{axelsson2022multimodal}
A.~Axelsson and G.~Skantze, ``Multimodal user feedback during adaptive robot-human presentations,'' \emph{Frontiers in Computer Science}, vol.~3, p. 741148, 2022.

\bibitem{macfie1989designs}
H.~J. MacFie, N.~Bratchell, K.~Greenhoff \emph{et~al.}, ``Designs to balance the effect of order of presentation and first-order carry-over effects in hall tests,'' \emph{J. Sensory Stud.}, vol.~4, no.~2, pp. 129--148, 1989.

\bibitem{hart1988development}
S.~G. Hart and L.~E. Staveland, ``Development of nasa-tlx,'' in \emph{Adv. in psych.}\hskip 1em plus 0.5em minus 0.4em\relax Elsevier, 1988, vol.~52, pp. 139--183.

\bibitem{glaser2017discovery}
B.~Glaser and A.~Strauss, \emph{Discovery of grounded theory: Strategies for qualitative research}.\hskip 1em plus 0.5em minus 0.4em\relax Routledge, 2017.

\bibitem{davis1989technology}
F.~D. Davis, R.~Bagozzi, and P.~Warshaw, ``Technology acceptance model,'' \emph{J Manag Sci}, vol.~35, no.~8, pp. 982--1003, 1989.

\bibitem{fujita2020customizing}
L.~Y. Fujita and S.~Y. Choi, ``Customizing physiologic alarms in the emergency department: a regression discontinuity, quality improvement study,'' \emph{J. of emergency nursing}, 2020.

\end{thebibliography}

\end{document}